# A Synergistic Approach to Digital Privacy

**Author:**
Christopher Gorog PMP, CISSP, CTO BlockFrame Inc., Founder International Alliance of Trust Chains, Founder Blockchain Development Community, Founder New Cyber Frontier, Published Author, Chair IEEE Blockchain Privacy & Security
https://www.linkedin.com/in/christopher-gorog/

**Editing Contributors:**
Michael Fargano, Independent System Engineer and Analyst, Technical Committee Chair Broadband Forum
https://www.linkedin.com/in/fargano

Zackary Foreman, Senior Software Engineer at CableLabs, Advanced Technology Group, Chair Association for Computing Machinery at UC Denver
https://www.linkedin.com/in/zackary-foreman-42708038/

Kyle Haefner, Lead Architect at CableLabs, Open Connectivity Foundation Chair Security Oversight Working Group https://www.linkedin.com/in/kyle-haefner/

Janelle Hsia, CIPM, CIPT, CIPP/US/E, CISA, GSLC, PMP, Founder and CEO of Privacy SWAN Consulting
https://www.linkedin.com/in/janellehsia/

Jason Rupe Ph.D., Principal Architect Cable Labs, Reliability, and Communications Societies, Co-Chair IEEE Blockchain Initiative, Managing Editor of IEEE Transactions on Reliability
https://www.linkedin.com/in/jrupe/

Nir Kshetri, Ph.D., Professor University of North Carolina at Greensboro, NIST Contributing Author, IEEE Reliability Society https://www.linkedin.com/in/nir-kshetri-1426a02/

Haobo Lai, Computer, Communications, Standards Association, Microwave Theory and Technology, Signal Processing, and Vehicular Technology Societies
https://www.linkedin.com/in/laihaobo/

Kent Lambert, COO BlockFrame Inc., President International Alliance of Trust Chains, Prior Colorado Senator, Sponsor of Privacy and Blockchain Legislation, IEEE Blockchain Privacy & Security, Co-Founder BlockChain Development Community
https://www.linkedin.com/in/kent-lambert-50861362/

Gurvirender Tejay Ph.D., Associate Professor University of Colorado Colorado Springs, Association for Computer Machinery (ACM), Association of Information Systems (AIS), International Federation for Information Processing (IFIP), Information Systems Audit and Control Association (ISACA), Information Systems Security Association (ISSA), Society for Information Management (SIM), International Association of Privacy Professionals (IAPP)   https://www.linkedin.com/in/gurvirender-tejay-344ab23/

Tim Weil, Security Feeds, IEEE Computer, Vehicular Technology, and Communications Societies, IEEE Blockchain Transactive Energy initiative. ICC 2024, Editor- IEEE IT Professional Magazine https://www.linkedin.com/in/tim-weil-a8b1952/

Gary Whitsett PMCISSP, Curriculum Development Lead FlatIron School/SecureSet, President  PhilosBDL Inc., IEEE Computer Society, Co-Founder BlockChain Development Community https://www.linkedin.com/in/garywhitsett/

*Abstract*- This paper outlines an approach for IEEE to take leadership for digital privacy to align many existing IEEE Societies and efforts in the areas of computer systems & applications security, organizational & global architectures, policy-supporting legislation, originating new standards, integrating compliance into technologies, and helping design decision-board infrastructures for governance bodies. Much of the current emphasis on evolving privacy technologies centers on big corporate enterprises and institutions, causing the industry to support corporate assets protection mainly. Fostering technology to empower individual privacy-enabling tools has lagged, and personal privacy has diminished because corporate big data applications have made sizable investments into exploiting private data. As one of the largest individual-member-based organizations, IEEE is urged to develop a collaborative approach for digital privacy with privacy-enabling technologies to benefit its members. The recommendations outlined define a prospective course that could result in future global individualized privacy capabilities which employ a combination of synergistic technologies such as distributed ledgers, differential privacy, homomorphic encryption, secure distributed multi-party computation, zero-trust architectures, proof-of-origin of data, software, or other techniques.  Such an effort would involve community engagement and outreach, academic peer-review events, the establishment of governance bodies, coordination & expansion of existing standards, and the development of publicly-accessible prototypes. Collaboration with other IEEE-sponsored efforts for transactive energy systems, confidentiality and security of healthcare records and devices, and other IEEE-funded projects will help magnify digital privacy investments already in progress in these applications of emerging technologies. An IEEE-led effort for Digital Privacy will provide global recognition, leadership, professional integrity, encouraging membership growth, and support a future technology capability for individualized privacy control.

# INTRODUCTION

Digital data breaches, stolen identities, cyber fraud, extortion, ransomware, destructive malware, and personal information abuse have elevated privacy to a significant societal issue, but solutions have been elusive. It is challenging to protect individuals' privacy rights when corporations, criminals, and even governments have incentives to collect and use personal or business data unethically or illegally [1]. A growing public outcry demands action and more-effective social, legal, and technological solutions, making privacy one of the most important topics of our age. This ubiquitous environment of insecurity presents IEEE with a unique opportunity to seek technical and policy solutions in perfect alignment with its core missions, Code of Ethics, and individual professionalism standards [2].

Technical solutions for the protection of corporate assets have often been resourced more favorably than individual privacy rights; however new privacy technologies, such as certain blockchain/distributed ledgers (BDL), have demonstrated exceptional value for distributed operations involving individualized control of digital assets [3], [4]. Future frameworks and platforms to apply evolving privacy technologies include distributed ledgers, differential privacy, homomorphic encryption, secure multi-party computation, and other techniques, which provide the exceptional potential to improve privacy, trust, and security [5], [6]. Digital Privacy will explore how to adopt new and evolving technologies to benefit individuals' and organizations' privacy and security to protect their data ownership. It will address the dichotomy between open public verification of data processing operations while simultaneously addressing the need to protect private or confidential data from a compromise of privacy. A Digital Privacy effort would support creating practical frameworks that empower individuals to protect the ownership of their digital data and technological solutions to secure individual privacy rights.

## A. Scope and Goals

The recommended Digital Privacy effort will support, adopt, and create frameworks and platform applications using technologies such as BDL to empower individual privacy and security. The Digital Privacy effort is recommended to be led by IEEE, which will strive to seek alliances with other global organizations, institutions, and individuals with similar objectives to achieve widely-accepted solutions and standards. The neutrality of operations and public visibility of controls will be maintained to the greatest extent possible for any sponsored or endorsed solutions by IEEE.

This IEEE effort will also seek to ensure that technologies, processes, and applications will satisfy compliance requirements such as the European Union (EU) General Data Protection Regulation (GDPR), the evolving Cybersecurity Maturity Model Certification (CMMC) framework, and best practices from leading industry methods[7]–[9]. The primary objective will be to provide privacy and security capabilities for individual citizens and organizations distributed across the digital world. The unique properties of new technologies, such as third-generation BDL, may integrate capabilities to prove the identities of people, hardware, and software, establish and prove trust, verify data, guarantee transaction integrity, and offer community incentives for collective participation for peer control and oversight [10]. Digital Privacy effort would strive to provide a roadmap with progressive steps to digital-age privacy goals achievable through technology.

## B. Ecosystem Scope

If effectively supported, IEEE Digital Privacy effort would eventually affect almost every person in the digital age. The priority to protect individual people's privacy has been recognized at the highest levels of society for the common good against those who exploit their vulnerabilities. The growth and innovation of "Big Data" have primarily been fueled by the proliferation and availability of personal data, usually without explicit consent. Massive accumulation of personal data and human behaviors are analyzed and used to predict, exploit, or manipulate user preferences, demographic and political trends, and industry insights that would not otherwise be possible. During the massive expansion of these processes, individuals' digital data and ownership rights have typically been ignored, underrepresented, and sometimes purposefully undermined, often with little oversight by government leaders. Private data exposure has also been a primary cause of identity theft and cybercrime against innocent victims.

IEEE professional standards support the concept that those who have the power to act on behalf of the good of others also have the responsibility to do so. The goal of a Digital Privacy effort would be to support the development and provision of tools and techniques that individuals and organizations need to reclaim control over their digital property rights. The effort will help create and apply digital privacy solutions to secure the confidentiality of data, allow data owners to control its authorized use, and verify its accuracy. Digital Privacy efforts will support designs for widely-adaptable and widely-scalable, modular, loosely-coupled, and open public interfaces applicable to a wide variety of use cases. Effectively applied, novel technologies could be a catalyst to disrupt negative trends and empower digital property rights to data owners.

Digital Privacy, as an IEEE-sponsored effort, should consider quality metrics of control, reliability, and management processes of any centralized platforms or infrastructure. Methods must maintain traceability and accountability for the use of data by other individuals and entities. Individuals desiring to protect their data from exploitation may incorporate privacy technologies that exhibit novel and positive emergent behaviors. IEEE Digital Privacy will identify and mitigate risks in architectures by neutrally-governing all components uniformly and fairly to maintain a highly-trusted ecosystem. Digital Privacy efforts will create, leverage, and support policies to maintain controls, oversight, integrity, and security for participant activities to meet these goals. IEEE's core purpose is to foster technological innovation and excellence for the benefit of humanity [11].

*C. Architectural Guidelines*

Leveraging the large community of IEEE Digital Privacy efforts can enlist professional participation and expertise from industries and other thought leaders to ensure best practices and techniques are considered from the earliest stages possible. Many novel technologies are still at a very early stage of maturity relative to their potential for growth. Satisfying only existing requirements or assuring compliance only with current standards might not adequately meet the goals this effort should address. Methodically identifying performance and non-performance requirements and defining new standards need validation from practical industrial insights for architectural design, test and evaluation, and compliance processes. When negative emergent behaviors exceed the abilities to control through best practices and expert guidance, the effort may examine and design new structural controls and tools that provide better options.

Effective system architectures need to consider data transitions from local control to cloud services and the use of new or evolving future technologies. Architectural designs should include designed-in solutions, backward compatibility with legacy components, and support for robust and flexible confidentiality frameworks. Technical architectures need to satisfy digital privacy rights, ethical transaction practices, and increase data handling accountability by restricting access to only authenticated and authorized users. Frameworks, architectures, and models need to consider distributed ledger, encryption, and other advanced technologies to protect data-at-rest and data-in-transit, independent of platforms or transmission media. Novel models and devices for handling confidentiality for aggregated data, and providing multiple degrees of separations to balance security with visibility and transparency, may be essential components within architectural designs for individually-owned data sets and public metadata search processes.

*D. Sustainable Funding Models*

Delivery of mass benefits to widely-distributed individuals must consider the economics of incentives for development, sustained operations, and secure components' maintenance and delivery. Distributed technologies such as cryptocurrencies have proved the ability to financially sustain their operational tempo, with incentives provided within the technology to reward both developers and users. Digital privacy enables technology using distributed ledger and methods such as tokenization of credits to incentivize projects and individuals economically while transitioning to self-sustainability. Investment partners, technology providers, and platform operators will be encouraged to engage with IEEE Digital Privacy efforts to provide valuable partnership opportunities, including access to tokenized stakes in supported technologies.

Ensuring participants' trust is a highly positive feature that will remain a strong focus of Digital Privacy and aligns with the IEEE Code of Ethics [2]. Contributing organizations and members will be professionally incentivized by aligning with this global, ethically-centered effort. As a largely individual-member-based organization, IEEE could improve the economic climate for many of its members by introducing global individualized privacy capabilities such as access to distributed ledgers and other privacy tools that allow a positive return on investment to its member base. Many supporting societies in IEEE recognize that addressing privacy and security on a global scale would positively impact its members while providing a highly innovative potential for society as a whole. IEEE has the opportunity to take a professional leadership role to engage with privacy and security-focused community organizations, local and global compliance organizations, and consumer protection and human rights advocates to ensure that proposed applications, frameworks, and solutions are as universally acceptable as possible. Interested societies and organizations that will benefit from the effort are welcome to provide initial funding or in-kind staff support.

Partner organizations will be encouraged to share information and knowledge gained from the effort and help create effective data aggregation and jurisdictional models. Ethical implementation should prohibit contributors from accessing private information by identifying authorized access points, operational sequences, and aggregated data components to protect identities, prevent unfair economic advantages by special interests, reduce conflicts of interest, and prevent public release of confidential information [12].

*E. Neutral Governance*

The governance of technologies to support individual privacy must be able to resist interests that could misuse such structures or to gain counterproductive advantages [13]. Any ambitious, wide-reaching effort using newly-designed features can also contain risks from hostile malicious activities. Efforts for Digital Privacy should support the principles that all policies, designs, processes, algorithms, and vetting operations should use open architectures, visibility, and transparency. The effort will strive to assimilate best practices for any adopted technology such as zero-trust-based architectures and distributed ledgers to display the integrity of operations and transactions while maintaining the privacy of encapsulated and encrypted data [14].

Open governance will include vetting participants, validation with partners and stakeholders, defining regional and legal jurisdictions, assessing the risks and trust of entities, maintaining fair incentives, restricting access to sensitive economic & financial aggregated data, reducing adverse cohesion, assigning audit operations and authorities, resolving disputes, and administering penalties for violations. The effort will enforce governance processes based on identifiable trust structures and algorithms using technologies to support quantifiable quality attributes and metrics [3].

### F. Collective Trust Among Participants

Features of evolving technologies such as blockchain/distributed ledgers, zero-trust architectures, multifactor-authentication, differential privacy, homomorphic encryption, and secure multi-party computation and proof-of-origin will increase the trust levels of transactions, the integrity of data, and the identification and participation of both human actors and system components [15]–[18]. The important social-physical bonds of building trust among people have been challenging to replicate in the technological era of digital electronic relationships leading to extreme social emergent behaviors such as disinformation [19]. Blockchains were initially applied to trustless monetary structures but have now shown exceptional promise in building trust among transacting parties.

As BDL moves beyond cryptocurrencies, the technologies' relationship-trust features maybe some of their most essential characteristics and contributions. Several novel and evolving methods of building and measuring trust will be incorporated into the Digital Privacy effort [3], including capabilities to support security, privacy, trust, data integrity, risk management, quality assessment, and abilities to identify malicious behaviors. Digital Privacy will commit to empowering individual trust in digital privacy rights, and minimize the effects of negative divergence.

### G. Change Management

Dynamic changes must be considered for any project and cannot be totally verified when a project starts. Change management for system and software upgrades, security patches, data sets, and changes in configuration, personnel, and standards are essential to maintain quality. IEEE Digital Privacy efforts should evaluate quantifiable trust-based approaches to track all change management transactions for operational components and project behaviors. Continuous Integration/Continuous Delivery (CI/CD) of digital changes is a best practice that should be leveraged for change management. Advantages of BDL for non-repudiable proof of transactions and permanent immutability of stored data offer unique and practical opportunities for automated and fully-accountable change management.

As changes occur, all components must be considered, entities engaged, and pre-existing relationships be provided with owner opt-out privileges presented in the GDPR [20]. Any changes requiring the addresses of members, existing ownership rights for access, visibility to aggregated data, and penalties for non-compliance of individuals or organizations must also be exercised in accordance with neutral governance policies and in accordance with the IEEE Code of Ethics.

### H. Standards and Compliance

Current IEEE efforts strive to define, improve, and introduce valuable standards for "distributed ledger"; however, to date, most of the focus has been on cryptocurrency-based blockchain implementations. In contrast, non-cryptocurrency, third-generation blockchains, thought more desirable, are still considered early-stage technology. The emphasis on digital privacy in part using BDL technology will provide advanced non-cryptocurrency capabilities, which will significantly enhance IEEE Blockchain standards efforts.

Many security standards have inadequate common definitions, older requirements, and methods that may not adequately describe or leverage the novel capabilities of advanced distributed ledgers, trust-chains, and other integrated privacy and encryption devices. The Digital Privacy effort will explore many of these areas and support IEEE standards with new definitions, standards, and compliance regimes that may be more appropriate in the current technological landscape.

Supporting an effort for Digital Privacy positions the IEEE to take a critical role to contribute to worldwide leadership for new privacy and security definitions and standardization. Included in privacy enabling technology, the abilities to include process quality and change management are vital features that are native and synergistic with privacy and forensic data linking. Privacy efforts would also ensure that the latest forensic tracking capabilities are considered for standardization so that internal processes can be quickly and accurately updated and promptly delivered worldwide. Globally-scaled third-generation blockchains could be particularly suitable to ensure accurate change management of standards-generated software, personnel requirements, privacy device configurations, accountability, oversight, and compliance with every change transaction.

IEEE has already focused on a few privacy standards. The following existing efforts will be incorporated and referenced into this Digital Privacy effort and expanded when appropriate.

IEEE P7002: Data Privacy Process [21]
https://standards.ieee.org/project/7002.html

IEEE P1912: Standard for Privacy and Security Framework for Consumer Wireless Devices [22]
https://standards.ieee.org/project/1912.html

IEEE P802E: Recommended Practice for Privacy Considerations for IEEE 802 Technologies [23]
https://1.ieee802.org/security/802e/

## I. *Program Management*

For the broadest effects, digital privacy prototypes should adopt requirements to manage rapid growth and globally-scalable security, privacy, and trust capabilities. Ideally, privacy-enabling technology should be designed into embedded devices and considered for all procedures, and thus any new designs would encompass the latest state of the practice. However, it is anticipated that many legacy systems will be supported, and hence a need for phased approaches to implement privacy technology will always exist. The architectures considered for data privacy efforts should consider a systems-of-systems approach to implement new components while leveraging bridging techniques from old practices. Such an approach will provide compatibility with existing legacy systems as much as possible, creating a gradual transition to re-designed upgrades or replacements as program managers consider them.

Identifying all possible requirements, features, and use cases are daunting and could lead to the failure of many efforts before they are stable and self-maintaining [24], [25]. Program management should build upon previous efforts and engage with known supporters and stakeholders. The approach of this effort is to leverage the management structure of ongoing projects which are at or near stability with a robust set of stakeholders [26], [27]. As the effort matures, more significant numbers of participants and stakeholders will be part of the management plan. Eventually, all processes could include public education and vetting by interested stakeholders on a global basis.

Program managers need to be aware of adverse risks from unfair manipulation by individuals or organizational stakeholders. The focus for each feature and overall project structures should support a modular system engineering approach for all processes, applications, and social behaviors. A strong focus on modularity will ensure that every structure, use case, or organization will be able to access common interfaces and repeatable open architecture designs equally.

IEEE Digital Privacy should make a conscious effort to minimize technologies that waste energy and natural resources through approaches such as cryptocurrency mining [28]. These challenges require new program management approaches and system controls. Prototypes initiated by the effort would strive to follow features of model-based code development and systems engineering principles to create replicable designs and reference architecture models that can be quickly and efficiently be re-used or quickly modified to apply to serve multiple use cases without the need to custom design each one[29]. Designs and architectures thus would focus on open architectures and modular loosely-coupled components, with the intent to streamline and accelerate the scalability of global privacy designs and devices and significantly reduce program timing and cost risks.

*J. Use Case Selection*

Independent projects, use cases, and participating business structures should be able to leverage the underlying components and models of prototypes. Some efforts may be performed and supported directly by IEEE, while other stakeholders might leverage open reference architectures produced as a result of the IEEE lead effort. It would be anticipated that IEEE-sponsored prototypes would be selected and run by IEEE project chairs and innovation leaders while being endorsed by IEEE societies. These prototype selections should be chosen to benefit each society by providing a usable application or framework outcome.

Selected IEEE-sponsored projects and use cases should be evaluated based on membership support and their adaptability to stakeholders' future industry adoption. IEEE currently recognizes three blockchain technology verticals in Healthcare and AI, Transactive Energy, and Security and Privacy [30]. By extension, other urgently-needed use cases, such as the integrity of transitions within critical supply chains, should also be examined. Digital Privacy efforts will align with and support these industry verticals and make these IEEE-sponsored projects a priority for implementation. For example, priorities will be given to protect consumer privacy for transactive bulk energy users, the security of energy transactions, and the confidentiality, privacy, and security of healthcare records, medical devices, and medical supply chains.

*K. Integration of Best Practices*

Some observers consider the privacy and security industry as immature, lagging behind the Internet's growth, and ineffective at meeting rapidly evolving requirements and growing threats. Comprehensive solutions are often elusive or very costly, and the rate of publicly-acknowledged data compromises and breaches continues to grow. Compliance requirements intended to reduce the occurrences or lessen the impacts of compromises are also often ineffective. Expert recommendations are often inconsistent and may change over time. In this landscape, security, trust, and privacy practices that currently appear normal, routine, or state of the art, might not remain indefinitely. As digital privacy technology has evolved, it has shown that current compliance requirements are not adequate, and future designs, operations, and standards must incorporate and adopt new approaches and best practices. Among those, requirements, tools, and procedures for individualized security and privacy must be a primary goal.

The balance between individualized control of owner-created data and the ability to use publicly-available metadata to search or index restricted content has been challenging to attain by current methods. Digital privacy technology such as distributed ledger has the unique ability to provide verifiable data proofs while maintaining the confidentiality of fully-encapsulated and encrypted data sets. When applying such techniques, other challenges arise if the visibility of data payloads is reduced. Implementation of new digital privacy technologies could redefine best practices and allow the identification and visibility of data quality, support confidentiality of aggregated data or economic trends, and verify the lineage of metadata, without the complete visibility or disclosure of confidential payloads. Also, effective technologies such as distributed ledgers to control the balances among individual anonymity, private data ownership, and the public verification and auditability of transactions could significantly support individualized privacy goals. The ability to publicly verify human actors and trust levels using repeatable trust verifications over time can revolutionize the ability to search for confidential data, control sovereign data instances, and support the protection and containment of data by their creators and owners.

*L. Conclusion*

The pretense of recommendations outlined in this approach would enable an effort to produce a global digital privacy capability for individuals. By supporting the outlined Digital Privacy efforts, IEEE would empower the creation of practical frameworks, technological solutions prototypes, and international governing structures to secure individual privacy rights. Since the technology to empower individual privacy-enabling-tools has lagged, personal privacy has suffered as corporate big-data applications come to market with sizable investments. With the rise in public awareness of privacy abuse compromise, the outcry for the protection of individuals continues to grow, making a prime opportunity for visibility of a concerted effort lead by IEEE to address this impending issue. As one of the largest individual-member-based organizations, IEEE is urged to develop a collaborative approach for digital privacy with privacy-enabling technologies to benefit its members. It is paramount to align the support of many societies in IEEE to ensure success and create an influence on a global scale to positively impact its members and the entire world. It is recommended that Digital Privacy becomes a top priority for IEEE and the worldwide community.

## ACKNOWLEDGMENT

This paper has been produced at the request of the IEEE Future Directions Committee as a response to a call for expertise on evolving privacy empowering technology. The use of this content is permitted by its author and contributing members for use by IEEE exclusively and is intended to be used only in and for publicly beneficial initiatives intended for the benefit of humanity.

This paper was developed through a collaborative process that brought together volunteers representing varied viewpoints and interests to achieve the final product. Volunteers are not necessarily members of IEEE and participated without compensation from IEEE.